\documentclass[a4paper]{jpconf}
\usepackage{graphicx}
\begin{document}
\title{3-D structures of planetary nebulae}

\author{W Steffen$^1$}

\address{$^1$ Instituto de Astronom\'{\i}a, Universidad Nacional Aut\'{o}noma de M\'{e}xico, Ensenada, B.C., M\'{e}xico\\}

\ead{$^1$ wsteffen@astro.unam.mx}

\begin{abstract}
Recent advances in the 3-D reconstruction of planetary nebulae are reviewed. We include not only results for 3-D reconstructions, but also the current techniques in terms of general methods and software. In order to obtain more accurate reconstructions, we suggest to extend the widely used assumption of homologous nebula expansion to map spectroscopically measured velocity to position along the line of sight.
\end{abstract}

\section{Introduction}

The main scientific motivation to generate 3-D representations of the structure of complex planetary nebula lies in the hope that they will contain information about the formation mechanism nebulae that is not obvious in the observations alone. A second reason for producing 3-D models is that of pure visualization, which helps to inspire new scientific insight or generate interest and better understanding by the general public.

Let me clarify what I mean by 3-D reconstruction by comparing the overall stages of 3-D research on planetary nebulae with that applied to study dinosaurs in palaeontology. The observations of a nebula correspond to extracting the remains of a dinosaur from the ground and the recording of the location and other properties of interest as they were found. More often than not, the relative location of the bones does not correspond to that in a living creature and many of the bones may actually be missing. Thus the observation and processing of the information from a palaeontological dig, corresponds to the observation of a nebula using a telescope and its instrumentation, followed by the processing of the data.

The 3-D reconstruction of the structure of a planetary nebulae based on imaging and spectro-kinematic information from the telescopic observations then corresponds to the interpretation of the data and the relative positioning of bones in the 3-D reconstruction of a skeleton. In both sciences, palaeontology and astronomy, the step from the collection of bones to the reconstructed skeleton and from processed observations to a 3-D reconstruction involve a significant amount of prior knowledge and assumptions on the type of creature and nebula that is being reconstructed.

Furthermore, the animal bones that were found or the radiation that was observed from a nebulae, are only a relatively small part of the complete being or nebula, respectively. Organs, muscles, skin and brain, among others, are what make up most of the living being, but these have been lost to decomposition. While any ordinary matter in space is expected to generate some kind of radiation, when we observe a nebula over a given spectral range with a limited sensitivity, a large part of the gas flow is likely to go undetected. The 3-D reconstruction of a nebulae therefore has to be ``fleshed out" in order to convey a complete picture of its present structure and potential past evolution. Multi-wavelength observations can help constrain the structure further even if a complete 3-D reconstruction of the multi-wavelength data is rarely possible, for the lack of a mapping to position along the line of sight.

The fact that astronomical observation directly only yields angular position on the sky is the key problem in 3-D reconstruction, since the structure along the line of sight for each pixel has to be found indirectly using additional constraints. Such constraints may have quite a diverse nature. In the case of an animal skeleton, the constraints may be the shape of the connecting regions between different bones. A tail bone simply will not sensibly fit to a vertebra near the neck. In the case of expanding nebulae symmetry is often a key constraining property, even in the case of non-spherical complex nebulae.

Wenger {\it et al.}~\cite{wag12} \cite{wlm13} developed automated numerical algorithms to reconstruct the 3D structure of nebulae using only an image with the assumption of general axi-symmetry of the emission. While their algorithm can handle minor deviations from axi-symmetry and exactly reproduces the projected structure for the original viewing direction, the ring-like artifacts typical for axi-symmetric models remain present. The algorithm works well when the symmetry axis is sufficiently inclined, but fails for small angles to the line of sight.

The key constraint for the 3-D reconstruction of more asymmetrical {\em expanding} nebula is the correlation between position and velocity. It was originally noted by Wilson \cite{w50} as a correlation between the expansion velocity and the nebula size for different spectral lines. This observation led to the general assumption of a homologous expansion law in which the position and velocity vectors are proportional to each other. The spectroscopically observed Doppler-velocity can then be mapped directly to the position along the line of sight, if the constant that links them can be determined from other constraints or observations \cite{stc04}.

Physically a homologous expansion can be obtain from a ballistic expansion, i.e. when the gas parcels are sorted by their constant velocity vector. This requires that there is no significant interaction with the external medium or later ejections in the form of a wind, jets or ``bullets". A rarefaction wave also generates a linearly increasing velocity field, but is not likely to be the origin of homologous expansion in planetary nebulae except in the quick ejection of dense massive stellar envelopes. A low-density wind interacting with dense clumps and filaments can also generate approximately homologous velocity fields for the clumps \cite{sl04}.

Some nebulae, such as NGC~6302, show no evidence from imaging, spectroscopy and internal proper motion measurements for significant deviations from a homologous expansion \cite{szw11}. The ejection mechanism for most of the optically visible nebula must therefore have been very short when compared to the evolutionary timescale.

Hydrodynamic simulations in one or more dimensions do, however, show that a homologous expansion law, in general, has only a limited applicability and may yield reconstructions with first order accuracy only \cite{sjs05} \cite{sgk09} \cite{s11}. There are two main deviations. First, from the two-wind shock interaction, along a given direction which shows discontinuous changes in velocity are crossed. Second, in regions where the flow is oblique to the shock surface, the flow direction changes discontinuously and can deviate significantly from the radial direction from the central star.

For bipolar and multipolar nebulae, there are {\em characteristic} deviations \cite{sgk09}, which have been confirmed by observations and their 3-D reconstructions in a study of an object that show an overall bipolar and axi-symmetric structure \cite{lgs12} \cite{stm14}. Such deviations occur strongest at mid-latitudes from the equatorial plane and generate false point-symmetric structures and flaring of the position-velocity (P-V) diagrams in these regions. Naturally, when jets are involved, interacting with earlier wind structures, strong local disturbances are expected in the form of a bowshock. The kinematic signatures of bowshocks from jets or bullets have been studied in detail in the context of young stellar objects and proto-planetary nebulae and do not follow a homologous expansion pattern \cite{rb86} \cite{bhf13}.

In general, in a pressure driven bubble, due to wind interaction or a blast wave, that evolves into a non-spherical structure, oblique shocks will appear. These shocks produce non-radial velocity fields, and hence deviations from homologous expansion in the kinematics.

We have to distinguish between the velocity field of the gas flow and the pattern speed of the visible structures. They may be different. This difference has a strong impact on the interpretation of internal proper motion measurement of the expansion from images taken at different times. In addition, the local ionization conditions may change and induce motion of a pattern that has no correspondence in the flow of the gas \cite{m04}.

All these deviations from the common assumption of homologous expansion for the reconstruction of the 3-D structure of planetary nebulae lead to the need of more detailed and systematic simulations, through (radiation-)hydrodynamical studies, with the specific aim of improving the methods for 3-D reconstruction of individual nebulae.
\begin{figure}[h]
\includegraphics[width=38pc]{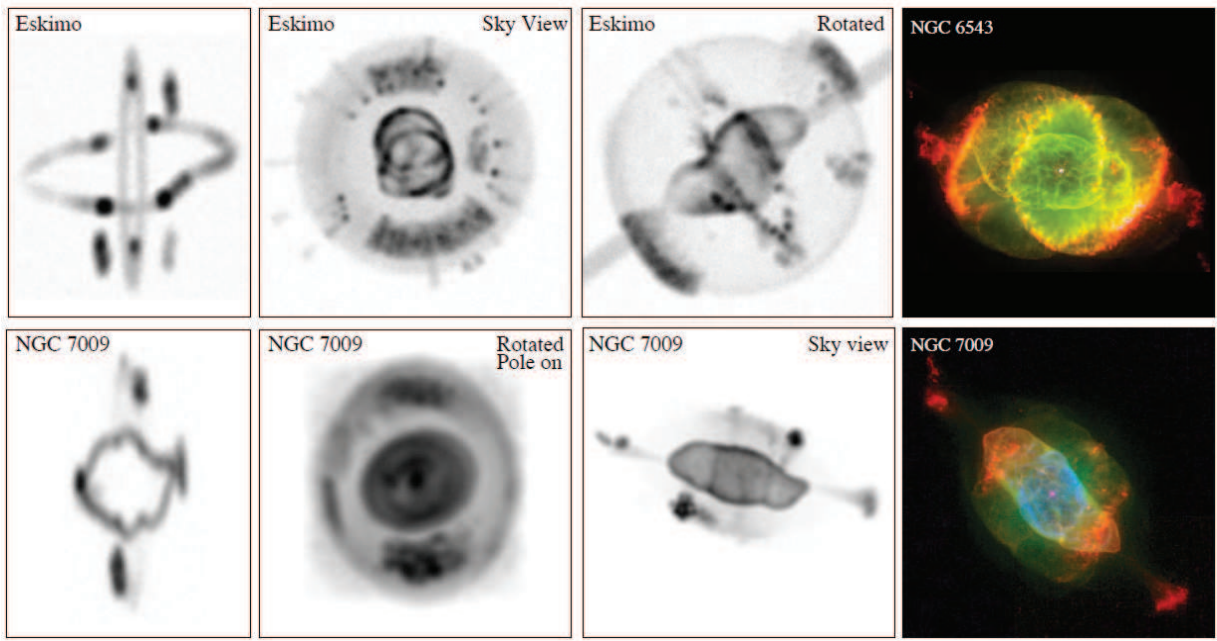}
\begin{minipage}[b]{38pc}\caption{\label{label}A comparison of images and P-V diagrams from three planetary nebulae seen from different real and synthetic viewing directions demonstrating their extreme similarities in structure and kinematics. Figure adopted from \cite{gls12}.}
\end{minipage}
\label{gls12.fig}
\end{figure}
\section*{Methods}

In recent years, in addition to the traditional custom programs, several publicly available software packages have been introduced to ease the 3-D modeling and reconstruction of astrophysical nebulae. Most of these packages specialize in particular aspects of the modeling, such as structure, photo-ionization, dust or molecular radiation transfer. One-dimensional photo-ionization has been successfully done for many years using CLOUDY \cite{fph13} and its extension to pseudo-3D with pyCloudy by C. Morisset \cite{gzm16}. The dimensional limitations of CLOUDY led Ercolano {\it et al.}~\cite{ebs03} to develop the fully three-dimensional Monte-Carlo radiation transfer code (MOCASSIN) that accepts any distribution of gas and dust density on a cartesian grid, including outputs from hydrodynamic simulations \cite{mf11}. Recently, Hubber {\it et al.}~\cite{hed16} have introduced a Voronoi grid algorithm in MOCASSIN, which is meant to improve spatial resolution at the same or reduced computational cost. In their paper the authors compare the original cartesian grid based algorithm with the Voronoi mesh method and find that the new method performs better on complex density distributions.

While using the radiation transfer computations as a post-processing step can not take into account the dynamical effects from the photon heating, it is a reasonable first approximation in highly supersonic flows, where the structure is dominated by wind-wind interaction shocks. Frank \& Mellema \cite{fm94} introduced methods to efficiently include photo-ionization from a central source, photon-heating and cooling in numerical hydrodynamical calculations. An example for the incorporation of diffuse radiation in 3-D hydrodynamic simulations is given in \cite{er13} for the case of Herbig-Haro outflows.

In the realm of molecular radiation transfer there are new publicly available options, too. The ARTIST/LIME combination is promising for modeling data of high resolution observations from the {\em Atacama Large Millimeter Array} (ALMA). A European team of researchers has been working on the development of the components for this software package in recent years \cite{bh10} \cite{pbg12}. This software was used, for instance, to model ALMA observations from a molecular 3-D spiral around the AGB star R Sculptoris, where the authors applied LIME radiation transfer to smoothed particle hydrodynamic simulations from the GADGET-2 code \cite{mmv12}.

CO molecular line transfer computations can now also be done in SHAPE \cite{sbk15}. SHAPE is a more general\-purpose 3-D interactive astrophysical laboratory, which has grown out of a basic interactive morpho-kinematic modeling program. The key difference to other modeling software is that the user does not interact with the code of the software. The setup, modeling and analysis of the models is done in an interactive user interface. In addition to 3-D mesh based interactive modeling, it incorporates a eulerian hydrodynamics module. The hydrodynamic simulations are also set up interactively in the same fashion as other morpho-kinematic models are set up using 3-D meshes. The interactive mesh setup in SHAPE allows very complex emission distributions to be modeled, which is the reason why most recent morpho-kinematic modeling in the area of planetary nebulae has been done using this software package.
\begin{figure}[h]
\label{cas15.fig}
\includegraphics[width=38pc]{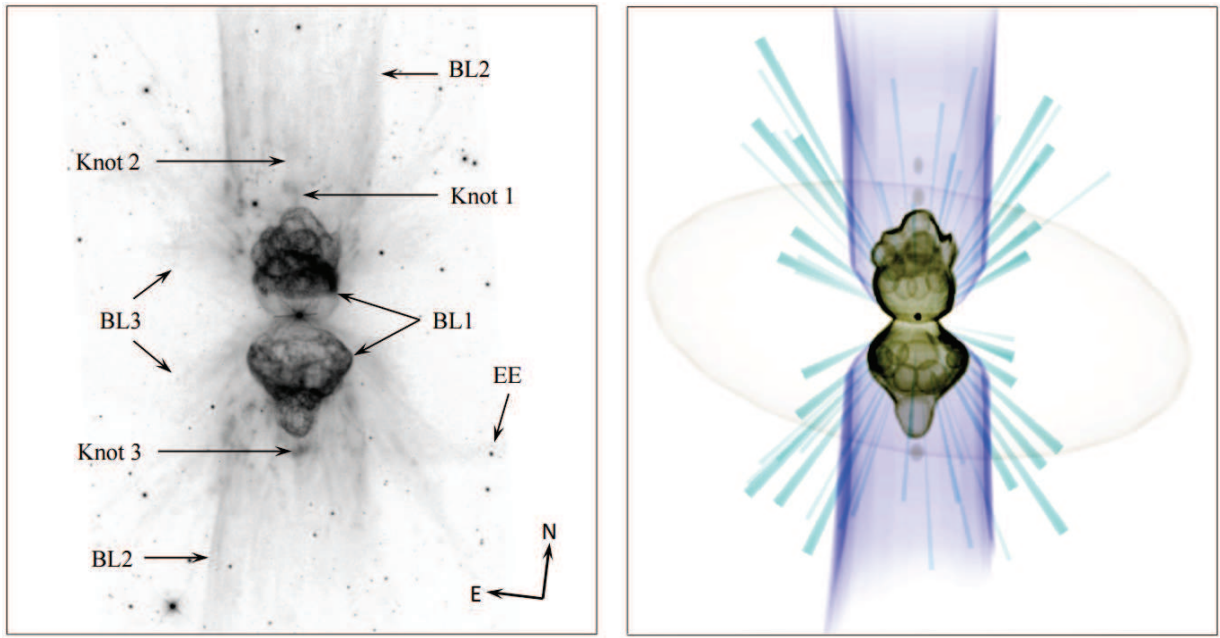}
\begin{minipage}[b]{38pc}\caption{\label{label}\cite{cas15} presented some of the most complex 3-D morpho-kinematic reconstructions to date. On the left the observed HST-image of Mz~3 is shown, while the right panel shows the 3-D reconstruction. Figure adopted from \cite{cas15}.}
\end{minipage}
\end{figure}
\section*{Some recent results}

One key property of planetary nebulae is their diversity in structure and ionization properties which manifest themselves in the variety that they show in their famous color images. The observed structures go from beautiful spherical bubbles, through highly collimated bipolar jet like nebulae to multipolar and even rather irregular shapes. Recent morpho-kinematic modeling has shown that, within each of the classes of nebulae, the diversity might not be as large as initially thought, which helps to reduce the number of formations mechanisms that have to be invoked for their explanation.

Chong {\it et al.}~\cite{cki12} and Hsia {\it et al.}~\cite{hcw14} showed that many images of planetary nebulae can reasonably be approximated by variants of a multipolar structure seen from different directions. Hence, projection effects heavily way in when classifying objects. Ambiguities introduced by projection must be eliminated by additional observational or theoretical constraints. The main tool to disentangle the ambiguities is of course the kinematics as obtained from spectroscopic observations of the gas velocity along the line of sight, with the caveats that have been described in the Introduction.

Using spectroscopic observations with good coverage of the kinematics in the Saturn Nebula (NGC 7009) and the Eskimo Nebula (NGC 2392), Garc\'{i}a-D\'{i}az {\it et al.}~\cite{gls12} showed that these two very complex bipolar nebulae have a 3-D emission structure that is very similar to that of the Cat´s Eye Nebula (NGC 6543)(Figure \ref{gls12.fig}). Using their reconstructions they have been able to adjust the viewing angle of each of the nebulae to allow a direct comparison of the images and P-V diagrams. The striking similarities lead them to conclude that the formation and evolution of these nebulae must have been basically the same. Following the analogy of reconstructing the skeleton of a dinosaur, now that the ``skeleton" of these nebulae has been obtained with the conclusion that they basically are of the same ``species", dynamical studies can be pursued to find out how they were ``born" and ``grew up".

A detailed study of the Homunculus Nebula around Eta Carinae revealed detailed structures that may be linked to the wind-wind interaction of the central binary star \cite{stm14} \cite{stm14m}. Two extrusions at a small angle from the equator, one on each of the bipolar lobes, are located in directions, with respect to the central binary, that corresponds very accurately to the opening angle and direction of the cone carved out by the interaction of the stellar winds as derived by Madura {\it et al.}~\cite{mgo12} using independent methods combining hydrodynamic simulations and spectroscopic observations of the central region of the nebula. Recent work by Teodoro {\it et al.}~\cite{tdh16} find a similar orientation based on spectral line intensities and width variations over several orbital cycles of the central binary.

The work on the Homunculus Nebula was the beginning of 3-D printing of astrophysical nebulae. Steffen {\it et al.}~\cite{stm14} and Madura {\it et al.}~\cite{mcg15} published 3-D mesh models of their results that have been 3-D printed. Such solid models are useful as a visual aid for scientific discussion or as a visualization tool for the general public.In particular, they help the visually impaired to literally getting a better ``grip" on the shape of objects in the universe, which are inaccessible to them through flat images. Hopefully 3-D printing becomes an increasingly useful tool for visualizing astrophysical findings, thereby reaching out for new audiences.

Imaging observations of planetary nebulae with the Hubble Space Telescope over more than a decade produced the hope to obtain detailed internal proper motion studies that would reveal kinematic information that is independent of spectroscopic measurements and serve as additional constraint to the velocity field beyond global expansion measurements (e.g. \cite{bgv12}). Szyszka {\it et al.}~\cite{szw11} performed a detailed proper motion analysis of NGC~6302. In the region that was measured, the nebula showed no significant systematic deviations from a homologous expansion within the dispersion of the measurements. A proper motion analysis of hydrodynamic simulations meant to reproduce the structure of NGC~6302 by Uscanga {\it et al.}~\cite{uve14} showed rather different results, which hardly any proper motion in the dense inner regions and increasing velocities further out, but with very strong scattering. Furthermore, the direction of the proper motion vectors as determined by their algorithm was rather erratic and no save conclusion can be obtained from these results. In part, this is likely to be due to the currently employed algorithms not properly capturing the motion of extended filamentary structures. Similar observational studies have been done by Garc\'{i}a-Diaz {\it et al.}~\cite{ggs15} for the Eskimo Nebula (NGC 2392), where the authors compare the proper motion obtained from two different methods.
\begin{figure}[h]
\includegraphics[width=28pc]{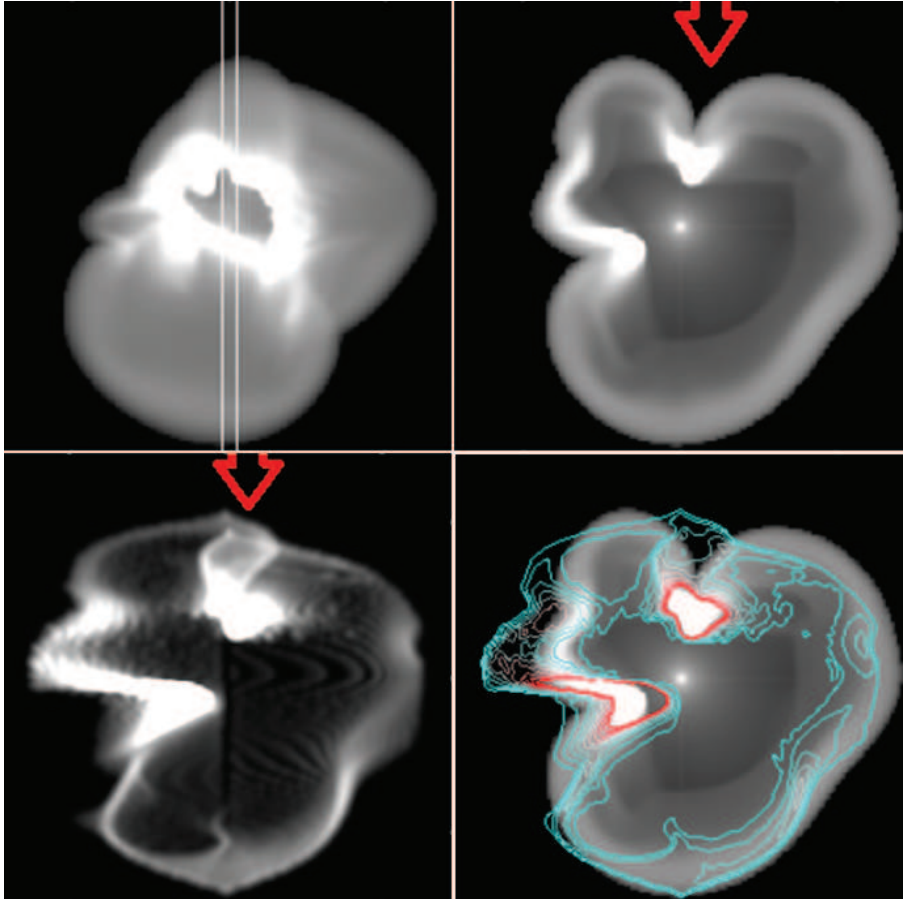}
\begin{minipage}[b]{38pc}\caption{Hydrodynamic simulations show that converging shock structures can produce topological defects in 3-D reconstructions. Top left: Density image of a simulated multipolar planetary nebula with synthetic spectrograph marked. Top right: spatial density cut at position of synthetic slit. Lower left: P-V diagram with kinematic feature of converging shocks marked. Lower right: superposition of spatial cut and P-V diagram (contours). Figure adapted from \cite{b15}.}
\label{b15.fig}
\end{minipage}
\end{figure}
Research on the quantification of proper motion in filamentary objects will be very helpful. Improvements along these lines could come from adapting pattern recognition methods to this problem such that they are capable of measuring the motion and distortions of knots, filaments or other recognizable features in easily quantifiable ways.

Most morpho-kinematic reconstruction work has concentrated on finding the 3-D structure of an object using an assumed homologous velocity field as a constraint. Observations and modeling techniques are now becoming sufficiently precise to allow, in some cases, to derive constraints on the velocity field itself, revealing deviations from the homologous expansion when assumptions on the symmetry can be made.

The observations and reconstruction of Hubble 5 by L\'{o}pez {\it et al.}~\cite{lgs12} showed that significant deviations from a homologous expansion velocity field had to be adopted to accommodate the apparent axi-symmetry of the double lobed planetary nebula.
The assumption of homologous expansion would have produced point-symmetries and deviations from axi-symmetry in the reconstruction, which would have been unreasonably well aligned with the line of sight. The derived deviations of the velocity field from homologous expansion were in agreement with the expectations from hydrodynamic simulations of wind-blown bipolar nebula shells. The deviations consisted in a non-linear increase of the magnitude of the velocity as a function of distance from the central star and a directional deviation that peaks at intermediate latitudes.

Also the paper by Clyne {\it et al.}~\cite{cas15} showed that a more flexible velocity field and appropriate assumptions on the structure of an object can lead to relevant information on details of the structure itself or on the velocity field. Their model for M~2-9 required a non-homologous expansion in order to maintain the apparent axi-symmetry. It turned out an additional expansion component away from the symmetry axis did the job, which is consistent with what is found in hydrodynamical simulations of such collimated structures.

Clyne {\it et al.}~\cite{cas15} also presented a reconstruction of Hen~2-104 (the ``Southern Crab"). This case is interesting since in addition to the global axi-symmetry, local assumptions on the ``allowed" regions along a well defined ring of the knots and filaments in the highly fragmented outer hourglass structure, led to the identification of two opposite knots on the symmetry axis within the hourglass. These might be slower residuals of the axial ejecta that are prominent further out or they may have been later ejections. In the same paper a reconstruction of Mz-3 is presented, which includes the small-scale ``bubbly" structures of the lobes (Figure \ref{cas15.fig}).

Hydrodynamical simulations analyzed by Berm\'{u}dez \cite{b15} revealed that converging curved shocks can produce topological artifacts in 3-D reconstructions, since the kinematic signatures in the P-V diagrams can be superimposed, while in the actual spatial distribution the shocks do not even get into contact yet (Figure \ref{b15.fig}).

\section{Summary}

In recent years the observational material for precise morpho-kinematic 3-D reconstruction of planetary nebulae has increased considerably, in particular with the optical {\it San Pedro M\'{a}rtir Kinematic Catalog of Planetary Nebulae} (L\'{o}pez, this volume) and with ALMA for molecular outflows. In addition, software developments for modeling complex 3-D structures (SHAPE), 3-D photo-ionization (MOCASSIN) and 3-D molecular radiation transfer (LIME-ARTIST, SHAPEMOL) has made it possible to achieve new insight. Similarities between nebulae that appear very different in their images have been found, thereby reducing somewhat the intrinsic variety of shapes. 3-D hydrodynamic simulations have also lead to new extensions \cite{ske13} to the already generalized interacting wind model originally devised by Kwok {\it et al.}~\cite{kpf78}. Studies have shown that the common assumption of homologous expansion for morpho-kinematic reconstructions can lead to misinterpretations and artificial structures, thereby jeopardizing the search for the formation mechanism. The detailed observational investigation of the kinematics together with hydrodynamic simulations, with global and local constraints on the structure, have allowed to show that at least in some planetary nebulae the expansion is non-homologous and information on the structure of the velocity field itself can be derived. Future detailed radiation-hydrodynamic characterizations of the velocity fields for the diverse structures of planetary nebulae should help to improve morpho-kinematic 3-D reconstructions.

\ack
The author acknowledges support by grant UNAM-PAPIIT 101014 and thanks the organizers and their institutions for support to attend the conference.

\end{document}